\documentclass[%
 reprint,
 amsmath,amssymb,
 aps,
prb,
showkeys
]{revtex4-2}

\usepackage{graphicx}% Include figure files
\usepackage{dcolumn}% Align table columns on decimal point
\usepackage{bm}% bold math
\usepackage{xcolor}% color fonts
\usepackage{multirow}
\usepackage{array}
\definecolor{red}{rgb}{1.0, 0.01, 0.24}

\begin{document}

%\preprint{APS/123-QED}

\title{Stacking-dependent magnetic ordering in bilayer ScI$_2$}%
%%% Force line breaks with \\

\author{Soumyajit Sarkar}
\email{dss.ph@brainwareuniversity.ac.in}
\affiliation{Department of Physics, Brainware University, Barasat, Kolkata-700125, India}%


\author{Soham Chandra}
\affiliation{Department of Physics, Brainware University, Barasat, Kolkata-700125, India}%
%  Second institution and/or address\\
%  This line break forced% with \\
% }%
% \affiliation{
%  Third institution, the second for Charlie Author
% \affiliation{%
%  Authors' institution and/or address\\
%  This line break forced with \textbackslash\textbackslash
% }%


\date{\today}% It is always \today, today,

%%%%%   Define new command
\newcommand{\Sc}{ScI$_2$}

\begin{abstract}
Stacking-dependent magnetism in two-dimensional van der Waals materials offers an effective route for controlling magnetic order without chemical modification. Here, we present a combined first-principles and finite-temperature study of magnetic ordering in bilayer ScI$_2$ with different stacking configurations. Using density functional theory with Hubbard-$U$ corrections, we investigate the structural, electronic, and magnetic properties of monolayer and bilayer ScI$_2$ in AA, AB, and BA stackings. The electronic structure exhibits a spin-polarized ground state dominated by Sc-$d$ states near the Fermi level. Mapping total energies onto an effective Heisenberg spin Hamiltonian reveals strong intralayer ferromagnetic exchange that is largely insensitive to stacking, while the interlayer exchange depends strongly on stacking geometry, favoring ferromagnetic coupling for AA and BA stackings and antiferromagnetic coupling for the AB stacking. Spin--orbit coupling calculations show that both monolayer and bilayer ScI$_2$ possess a robust out-of-plane magnetic easy axis. Finite-temperature Monte Carlo simulations indicate that all bilayer configurations sustain magnetic ordering at and above room temperature, with ordering temperatures in the range 360--375~K, as confirmed by Binder cumulant analysis and finite-size scaling. These results demonstrate that stacking geometry enables control of the magnetic ground state in bilayer ScI$_2$ without significantly affecting its thermal stability.
\end{abstract}

%Use showkeys class option if keyword
\maketitle

%\tableofcontents
\section{\label{Sec:Intro} Introduction}
The discovery of intrinsic magnetism in two-dimensional (2D) van der Waals (vdW) materials has revolutionized the field of condensed matter physics, 
offering unprecedented opportunities for exploring low-dimensional quantum phenomena and developing next-generation spintronic 
devices~\cite{burch2018magnetism,Blei2021,Bedoya-pinto2021Science,Liu2023}. 
These atomically thin magnets, such as CrI$_3$ and Fe$_3$GeTe$_2$, exhibit long-range magnetic order down to the 
monolayer limit, challenging traditional theories like the Mermin-Wagner theorem by incorporating magnetic anisotropy to stabilize ordering against 
thermal fluctuations\cite{burch2018magnetism}. Layer-dependent magnetic properties are prominent in these systems, as evidenced by 
the transition from intralayer ferromagnetic (FM) ordering in monolayers to interlayer antiferromagnetic (AFM) coupling in CrI$_3$ 
bilayers\cite{Sivadas2018NanoLett,Jiang2019PRB} and enhanced interlayer FM exchange in Fe$_3$GeTe$_2$ with decreasing layer thickness\cite{park2022layer}. In 
multilayer structures, the interlayer magnetic coupling—whether FM or AFM—plays a crucial role in determining the overall magnetic behavior, with 
applications in magnetic tunneling junctions, memory devices, and quantum computing\cite{Wang2024PRB}.

A key challenge in 2D magnetism is the precise control of interlayer interactions without invasive methods like doping or external fields~\cite{Li2024NatCommun,Soenen2023}.
Recent studies have demonstrated that stacking configurations, including sliding and rotation, can profoundly influence interlayer exchange coupling in vdW 
bilayers\cite{jiang2021recent}. For instance, in CrBr$_3$ bilayers, direct observations via scanning tunneling microscopy 
revealed stacking-dependent 
transitions between FM and AFM states, attributed to variations in superexchange pathways mediated by halide anions\cite{song2019direct,Yang2023JACS,Li2024NatCommun}. 
Similarly, in CrI$_2$ systems, theoretical investigations have shown that reversed stacking can induce altermagnetism or ferroelectric properties, 
highlighting the sensitivity of magnetic order to interlayer registry\cite{Sun2024PRB}. A recent theoretical study further illustrates how interlayer 
exchange coupling can drive magnetic phase transitions between ferromagnetism and altermagnetism in 2D lattices\cite{ga2025interlayer}. These findings 
emphasis the role of superexchange interactions, where orbital hopping between magnetic ions via non-magnetic ligands dictates the sign and strength 
of magnetic coupling, as described by the Goodenough-Kanamori rules\cite{goodenough1955theory,kanamori1959superexchange}.\\
Despite these advances, many 2D magnetic materials remain underexplored, particularly those with unconventional 
stoichiometries like AB$_2$-type structures\cite{nomura2021structure,Fukuda2021}. 
Scandium diiodide (ScI$_2$), a predicted 2D material with intrinsic FM ordering within each layer, represents a 
promising candidate for tunable magnetism due to its vdW nature and potential for structural manipulation\cite{acosta2022machine}.
However, the interlayer magnetic behavior in bilayer ScI$_2$, especially under varying stacking modes, has not been systematically investigated. 
Understanding how sliding and rotation alter superexchange mechanisms could unlock new strategies for mechanical or electrostatic control of magnetic states.\\ 
In this work, we employ DFT-based first-principles calculations combined with Monte Carlo simulations to study the 
stacking-dependent magnetic ordering in bilayer ScI$_2$ systematically. Motivated by the growing 
interest in tunable two-dimensional van der Waals 
magnets for spintronic applications, we focus on how different stacking registries influence the interlayer magnetic coupling and 
thermal stability of the system. Our results reveal that AA stacking stabilizes FM interlayer coupling, whereas lateral translation to 
AB/BA stackings enables reversible switching between FM and AFM states, driven by changes in orbital hybridization 
and exchange pathways. We further analyze the corresponding exchange interactions and magnetic transition temperatures, highlighting 
the strong sensitivity of magnetic behavior to stacking geometry. These findings provide deeper insight into the controllable magnetism 
of ScI$_2$ bilayers and underline their potential for designing stacking-engineered two-dimensional magnetic devices.\\
The manuscript is organized as follows. First, we describe the computational methodology employed in this work. This is followed by a detailed presentation of the structural, electronic, and magnetic properties, together with a discussion of the microscopic origin of the stacking-dependent magnetic behavior. Finally, we summarize the main conclusions and discuss their broader implications for two-dimensional magnetism and future spintronic applications.
%%%%%%%%%%   SECTION  %%%%%%%%
\section{\label{Sec:Method}Computational Details}
All first-principles calculations in this study were performed using the plane-wave pseudopotential method based on density functional theory (DFT) as
implemented in the Quantum ESPRESSO package~\cite{Giannozzi2009, Giannozzi2017}. The exchange-correlation interaction was treated within the framework
of the generalized gradient approximation (GGA) using the Perdew–Burke–Ernzerhof (PBE) functional~\cite{Perdew1996}. 

The projector augmented-wave (PAW) method~\cite{Blochl1994} was employed to describe the electron-ion interactions, 
with pseudopotentials taken from the standard Quantum ESPRESSO pseudopotential library. 
%A kinetic energy cutoff of \textbf{XXX}~Ry for the plane-wave basis set and \textbf{YYY}~Ry for the charge density was used, ensuring convergence of total energy and forces.
The total energies are relaxed to $3 \times 10^{-7}$ eV/cell and crystal structures (all degrees of
freedom, atomic positions, and cell) until forces on ions are below 0.005 eV\AA.
Brillouin zone integration was carried out using a Monkhorst–Pack $13 \times 13 \times 1$ $k$-point grid~\cite{Monkhorst1976}. 
A vacuum layer of at least $15$\AA\ was applied along the direction out-of-the-plane to avoid spurious interactions between periodic 
images of the ScI$_2$ monolayer or bilayers. 
Systematic convergence tests for the plane-wave cutoff energy and Monkhorst–Pack $k$-point 
mesh were performed to ensure the accuracy and reliability of the calculated results.
To properly describe the localized Sc~$3d$ electrons, we employed the DFT+$U$ 
method in the Dudarev formulation\cite{Dudarev1998}, in which only the effective 
parameter $U_{\mathrm{eff}} = U - J$ is required. An on-site Coulomb correction of 
$U_{\mathrm{eff}} = 1.7$~eV was applied to the Sc~$2d$ states. This choice is 
consistent with earlier first-principles studies on Sc-based halide monolayers 
and related two-dimensional transition-metal systems, where 
$U_{\mathrm{eff}}$ values in the range of $2$~eV have been found to correctly 
capture the electronic and magnetic properties\cite{Wu2023}.  
Long-range dispersion interactions were accounted for using the Grimme 
DFT-D3\cite{Grimme2010} van der Waals correction as implemented in \textsc{Quantum~ESPRESSO}.

The thermodynamic stability of the \Sc\ monolayer was assessed by computing the formation energy
with respect to its constituent elements. 
We performed classical Monte Carlo simulations to investigate the thermo-magnetic properties of bilayer $ScI_{2}$ configurations.
The simulations were carried out on square computational grid of linear dimensions 
$L = 30, 40, 50, 60, 70, 80, 100, 120, 140, 160$ and $200$, 
with periodic boundary conditions applied along the in-plane directions.
Although bilayer ScI$_2$ possesses an intrinsic hexagonal lattice geometry, the Monte Carlo
simulations were implemented on a square 
computational grid by selectively assigning the appropriate nearest- and next-nearest-neighbor interactions to preserve the original 
hexagonal coordination and exchange topology, following the standard mapping procedure commonly used in spin-lattice simulations\cite{Chandra2021JPCS}.
Spin configurations were updated using the local Metropolis algorithm,
where trial moves are accepted with probability, $p = \min \left\{ 1, \exp\left( -\frac{\Delta E}{k_{B}T} \right) \right\}$.
At each temperature, the system was equilibrated for $5\times10^{4}$ Monte Carlo sweeps per spin (MCSS), followed by an additional $5\times10^{4}$ MCSS over which thermal averages were computed. 
To improve statistical accuracy, all reported quantities were further averaged over $10$ microscopically distinct but macroscopically equivalent realizations. The sublayered magnetization i.e. magnetization per spin of the individual layers, is defined as:
\begin{eqnarray}
m_{top} = \left\langle \frac{1}{N_{Sc}} \sum_{i \in top} S_i \right\rangle \\
m_{bot} = \left\langle \frac{1}{N_{Sc}} \sum_{i \in bot} S_i \right\rangle 
\end{eqnarray}
For the ferromagnetic configurations the total magnetization, $m_{t}$, is the order parameter, and is defined as:
\begin{equation}
	m_{t} = \frac{1}{2} \left| m_{top} + m_{bot} \right| 
\end{equation}
For the antiferromagnetic configuration the staggered magnetization, $m_{s}$, is the order parameter, and is defined as \cite{Matsuura1976:JPSJ}:
\begin{equation}
m_{s} = \frac{1}{2} \left| m_{top} - m_{bot} \right| 
\end{equation}
and the magnetic susceptibility is calculated by:
\begin{equation}
\chi_{m} = \frac{1}{k_{B}T}
\left(
\langle m_{t,s}^{2} \rangle - \langle m_{t,s} \rangle^{2}
\right),
\end{equation}
whose peak provides a preliminary estimate of the magnetic transition temperature. To obtain a size-independent and more precise determination of the critical temperature, we compute the fourth-order Binder cumulant
\begin{equation}
U_L = 1 - \frac{\langle m_{t,s}^{4} \rangle}{3 \langle m_{t,s}^{2} \rangle^{2}} ,
\end{equation}
for all system sizes and identify the transition temperature from the crossing points of $U_L(T)$ curves for successive system-size pairs, $(L,2L)$. The analysis is restricted to the temperature window $355\,\mathrm{K} \leq T \leq 385\,\mathrm{K}$, and only the first crossing within this range is retained to avoid spurious numerical artifacts and finite-size effects.

\section{Results and Discussions}\label{sec:Result}
\subsection{Structural Properties of Monolayer and Bilayer \Sc}
\subsubsection*{Monolayer Geometry}

Monolayer scandium diiodide (ScI$_2$) can crystallize in two possible polymorphs distinguished by the local 
coordination environment of the Sc atoms—
namely, the \textit{octahedral} (1T) and \textit{trigonal prismatic} (1H) structures. Both are derived from the CdI$_2$-type layered structure, common 
among transition-metal dihalides, and consist of Sc layers sandwiched between two iodine planes forming an I–Sc–I trilayer.

In the 1T phase, Sc atoms are octahedrally coordinated by six I atoms, forming edge-sharing ScI$_{6}$ octahedra arranged in a hexagonal 
lattice with space group $P\overline{3}m1$. This configuration is analogous to that found in monolayer CrI$_2$~\cite{Jiang2019PRB}. 
In contrast, the 1H phase exhibits trigonal prismatic coordination of Sc by I atoms (space group $P6_{3}/mmc$), similar to that of MoS$_2$ and other transition-metal dichalcogenides~\cite{Xu2016AIP}. 
The two phases differ primarily in the stacking of I–Sc–I layers and the symmetry of the 
coordination environment.
Our total energy calculations indicate that the 1T phase of ScI$_2$ is energetically more favorable than the 1H phase by approximately 
$\Delta E \approx 0.11$~eV per formula unit. 
This energetic preference arises from the smaller crystal-field splitting and better overlap between 
Sc~3\textit{d} and I~5\textit{p} orbitals in the octahedral coordination, leading to a more stable electronic configuration\cite{kanamori1959superexchange}.
The optimized lattice constant of the 1T monolayer is about $a = 3.9$~\AA, and the Sc–I bond length is $2.8$~\AA.
For comparison, the 1H phase relaxes to a slightly larger lattice constant due to reduced in-plane orbital overlap.

The variety of stacking registries provides a natural platform to modulate interlayer orbital overlap and superexchange pathways between adjacent Sc layers. 
These geometric variations, as will be shown in the following subsections, lead to significant changes in the interlayer magnetic 
coupling, allowing the system to switch between ferromagnetic (FM) and antiferromagnetic (AFM) alignments through controlled stacking translations or rotations.

\subsubsection*{Bilayer Stacking Configurations}

\begin{figure}[ht]
    \centering
    \includegraphics[width=0.9\linewidth]{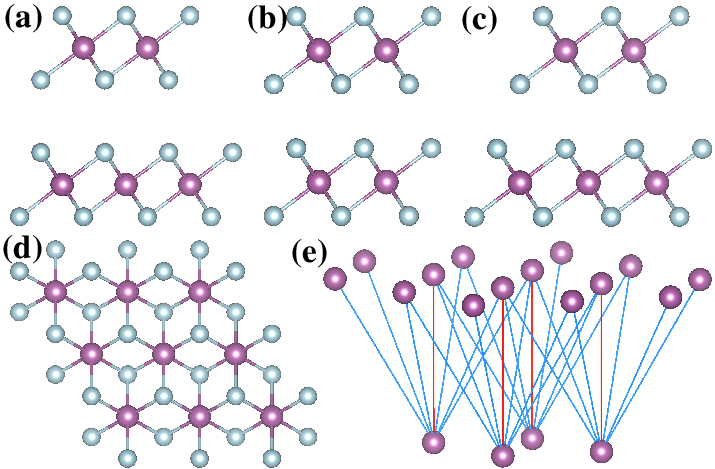}
    \caption{Side view of bilayer ScI$_2$ showing three distinct stacking configurations: (a)BA, (b) AA, and (c)AB. (d) Top view of monolayer ScI$_2$. Scandium and iodine atoms are represented in violet and grey, respectively. (e) Schematic diagram showing first-nearest neighbor (red line) and second-nearest neighbor (blue line) inter-layer interactions in bilayer ScI$_2$. Iodine atoms are omitted for clarity of the interaction path.}
    \label{fig:stacking}
\end{figure}

The bilayer ScI$_2$ structures were built using the energetically favorable 1T phase as the building block.
We consider three different stacking configurations based on the ground-state 1T monolayer structure: 
AA, AB, and BA. These are generated through combinations of lateral translation of vertical stackings, as illustrated in Fig.~\ref{fig:stacking}.

The AA stacking is formed by directly placing one monolayer atop the other with perfect registry of atomic positions, 
preserving the out-of-plane mirror symmetry $M_z$ in the $xy$-plane. The AB and BA configurations are obtained from AA by laterally translating the upper layer by fractional vectors 
$\mathbf{t}_1 = (-1/3, -2/3)$ and $\mathbf{t}_2 = (1/3, 2/3)$ in units of the primitive lattice vectors, respectively. Note that the BA stacking is 
equivalently obtained by a 180$^\circ$ rotation of the AB configuration about the $y$-axis.\\
The optimized interlayer distances $d_{\perp}$ and stacking energies $E_{\text{stack}}$ (defined as $E_{\text{stack}} = E_{\text{bilayer}} - 
2E_{\text{monolayer}}$ per formula unit) are summarized in Table \ref{tab:stacking}. The AA stacking exhibits the largest interlayer separation 
($d_{\perp} \approx 3.75$ \AA), consistent with weak van der Waals binding and minimal orbital overlap in eclipsed iodine positions. In contrast, 
translational shifts in AB and BA reduce $d_{\perp}$ to $\sim 3.45$ \AA, indicating enhanced interlayer cohesion due to staggered iodine alignment.\\
Among the three configurations, AB configuration emerges as the lowest-energy structure, with $E_{\text{stack}} = -0.22$ eV/f.u.
and BA configuration has a comparable energy. The AA configuration has a little higher energy $-0.18$ eV/f.u.
These structural variations lay the foundation for the observed diversity in interlayer magnetic coupling, 
as discussed in subsequent sections.
To further verify the stability of the proposed bilayer ScI$_2$ structures, we performed phonon dispersion calculations for all 
considered stacking configurations using the finite displacement method. The absence of imaginary phonon frequencies 
throughout the Brillouin zone confirms their dynamical stability and supports their potential experimental feasibility. A detailed 
discussion on the calculation of the phonon spectra is provided in the Supplementary Information (Fig.~S3).

\begin{table}[b]
\caption{\label{tab:stacking} Optimized interlayer distance $d_{\perp}$ and stacking energy $E_{\text{stack}}$ (per formula unit) for different bilayer configurations of ScI$_2$.}
\begin{ruledtabular}
\begin{tabular}{lcc}
Configuration & $d_{\perp}$ (\AA) & $E_{\text{stack}}$ (eV/f.u.) \\
\hline
AA           & $3.75$ & $-0.18$ \\
AB           & $3.40$ & $-0.22$ \\
BA           & $3.45$ & $-0.21$
\end{tabular}
\end{ruledtabular}
\end{table}

%%%%%%%%%%%%   FIGURE STRUCTURE  %%%%%%%%%%%

\subsection{Electronic structure of monolayer and bilayer ScI$_2$}

\begin{figure}[t]
    \centering
    \includegraphics[width=0.9\linewidth]{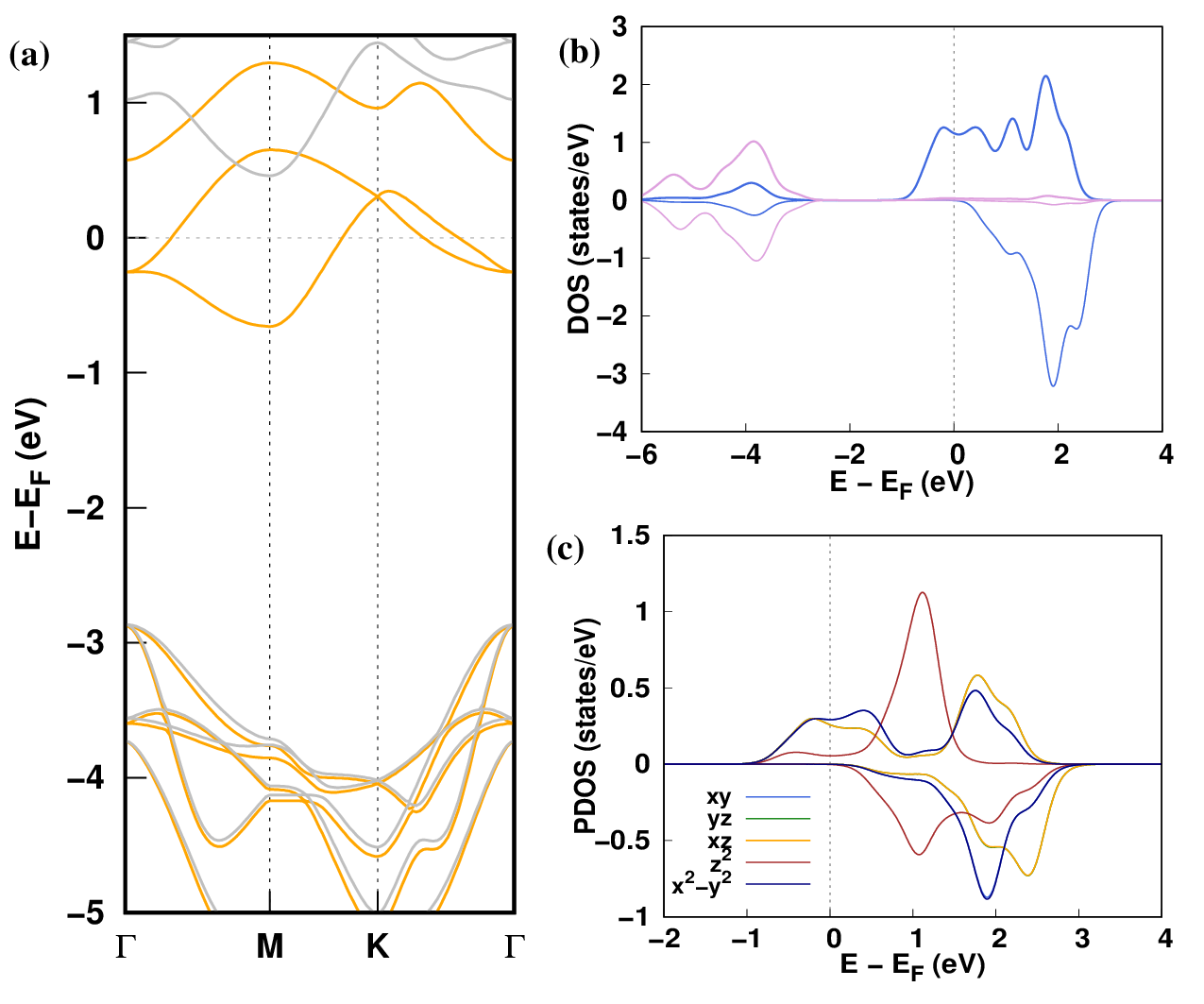}
    \caption{Electronic structure of monolayer ScI$_2$. 
    (a) Spin-polarized band structure along high-symmetry directions of the Brillouin zone. 
    (b) Spin-polarized density of states showing contributions from Sc-$d$ and I-$p$ orbitals. 
    (c) Spin-polarized density of states projected on the five $d$ orbitals of Sc.
    The Fermi level $\left( E_F\right)$ is set to zero of the energy axis.}
    \label{fig:mono_ele}
\end{figure}

Following the structural optimization of monolayer and bilayer \Sc~in different stacking configurations, we now analyze their electronic
structures to establish a reference for the stacking-dependent magnetic interactions discussed in subsequent sections. 
Electronic band structures and densities of states (DOS) were calculated for the optimized geometries, with emphasis on qualitative trends in orbital character and stacking-induced modifications in the electronic structure.
The monolayer \Sc~exhibits a spin-polarized electronic structure dominated by Sc-$d$ states near the Fermi level, 
as shown in  Fig.~\ref{fig:mono_ele}.
Owing to the nominal Sc$^{2+}$ valence state with a single $d$ electron, only one majority-spin band crosses the Fermi level, while the minority-spin channel remains gapped. This electronic configuration clearly indicates the half-metallic nature of monolayer \Sc.
The projected DOS further confirms that the electronic states in the immediate vicinity of the Fermi energy $(E_F)$ 
arise predominantly from Sc-$d$ orbitals, whereas the I-$p$ states are located at significantly lower energies, 
approximately $4$~eV below $(E_F)$, and exhibit appreciable hybridization with the Sc-$d$ states.

Such narrow $d$-derived bands close to the Fermi level are a common feature of two-dimensional transition-metal magnetic systems and reflect the reduced dimensionality and relatively localized nature of the transition-metal $d$ electrons \cite{Gong2017Discovery,Huang2017Nature, burch2018magnetism}. The calculated band structure of monolayer ScI$_2$ shows moderately dispersive bands, consistent with this picture and with previous first-principles studies on related two-dimensional magnetic halides\cite{Sarkar2020PhysRevMaterials,Sarkar2021PhysRevB}.

%%%%%%%%%%%  Figure Monolayer %%%%%%%%
\begin{figure}[t]
    \centering
    \includegraphics[width=0.9\linewidth]{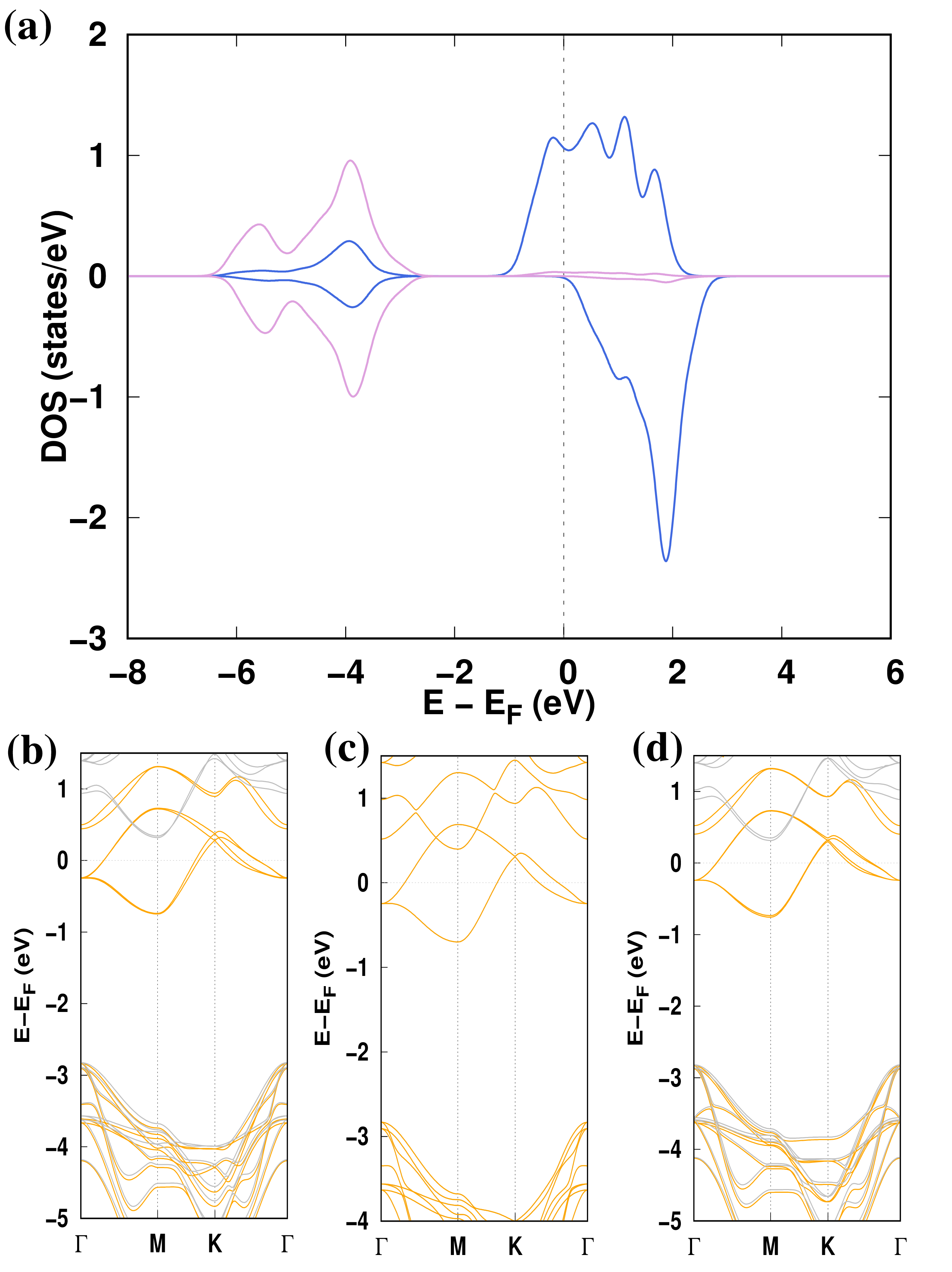}
    \caption{Stacking-dependent electronic structure of bilayer ScI$_2$. 
    (a) Projected density of states for the AA-stacked bilayer, highlighting Sc-$d$ and I-$p$ orbital contributions. 
    (b) Spin-polarized band structures for AA, AB, and BA stacking configurations. Orange (grey) bands present the majority (minority) spin channel.
    The Fermi level is set to zero on the energy axis.}
    \label{fig:bi_ele}
\end{figure}

%%%%%%%%%%%%%%%%%%%%%%%%%%%%%%
When two ScI$_2$ layers are combined to form bilayers, the overall electronic structure remains qualitatively similar to that of the 
monolayer; however, clear stacking-dependent modifications emerge. While the intralayer electronic features are largely preserved, the 
relative lateral displacement between the layers in AA, AB, and BA stackings alters the degree of interlayer orbital overlap. 
Fig.~\ref{fig:bi_ele} (lower panel) shows that 
subtle changes in band dispersion near the Fermi level, particularly for bands with dominant Sc-$d$ character. 
These variations indicate stacking-sensitive interlayer hybridization without leading to a drastic reconstruction of the 
electronic structure, confirming that the bilayers retain their quasi-two-dimensional nature.

A closer inspection of the projected DOS in Fig.~\ref{fig:bi_ele}(a) shows that stacking-induced changes primarily 
affect the Sc-$d$ states, which play a central role in determining the magnetic exchange interactions. 
The I-$p$ states continue to act as mediators of hybridization, 
forming Sc--I--Sc pathways both within and across layers. 
Depending on the stacking configuration, the relative alignment of Sc and I atoms across the interface modifies these hybridization 
channels, leading to stacking-dependent electronic fingerprints near E$_F$. Although these differences are modest in the 
electronic spectra, they have important consequences for the sign and magnitude of interlayer magnetic exchange interactions, as 
discussed in the following section.
%%%%%%%%%  Figure DOS & Band Bilayer %%%%%%%%

It should be emphasized that the present analysis does not aim at bandgap engineering or precise determination of electronic gaps. In 
low-dimensional and correlated systems, bandgap values are known to be sensitive to the choice of exchange--correlation functional, 
the Hubbard $U$ parameter\cite{Sarkar2020PhysRevMaterials}, and many-body corrections beyond standard density functional theory \cite{Qiu2013PRL, Ramasub2012PRB, Hybersten1986PRB}. 
Previous studies employing advanced approaches such as G$W$ and related methods have demonstrated substantial 
quantitative corrections to DFT band gaps in two-dimensional materials \cite{Qiu2013PRL, Ramasub2012PRB}. In this context, the present 
focus on robust qualitative trends in orbital character and stacking-dependent electronic modifications provides a physically 
transparent and reliable basis for understanding the magnetic behavior of bilayer ScI$_2$.

\subsection{Magnetic exchange interactions}

The stacking-dependent magnetic behavior of bilayer ScI$_2$ is governed by the nature and strength of magnetic exchange interactions within and between the layers. To quantify these interactions, total energies obtained from first-principles calculations were mapped onto an effective Heisenberg spin Hamiltonian of the form
\begin{equation}
H = - \sum_{\langle ij \rangle} J_{ij}\, \mathbf{S}_i \cdot \mathbf{S}_j ,
\end{equation}
where $J_{ij}$ denotes the magnetic exchange coupling between spins $\mathbf{S}_i$ and $\mathbf{S}_j$, and the summation runs over 
relevant intra- and interlayer neighbor pairs. Within this convention, positive (negative) values of $J_{ij}$ correspond to 
ferromagnetic (antiferromagnetic) exchange interactions.
For bilayer ScI$_2$, we considered the dominant intralayer first nearest-neighbor exchange interaction ($J_{\parallel}$) together with 
the first and second nearest-neighbor interlayer exchange interactions ($J_{1}$ and $J_{2}$), which primarily determine the magnetic 
ground-state ordering. 
The $J_1$ and $J_2$ exchange interactions were estimated using the shortest Sc–Sc bond distances and the dominant I-mediated 
superexchange pathways in the bilayer structure. Since longer-range interactions are significantly weaker due to the increased atomic 
separation and reduced orbital overlap, exchange terms beyond the second nearest neighbors were neglected in the effective Heisenberg 
Hamiltonian. The corresponding exchange paths and bond lengths are illustrated in Fig.~\Ref{fig:stacking}(e).
The exchange parameters were extracted using an energy-mapping approach based on the total energies of several 
ordered collinear spin configurations, including the FM state and different AFM arrangements 
specifically considered to isolate the individual exchange pathways. These configurations include one intralayer AFM state and two 
distinct interlayer AFM states corresponding to the first and second nearest-neighbor interlayer couplings. A detailed description of 
the magnetic configurations and the corresponding energy-mapping procedure is provided in the Supplementary Information (Sec.~S4 and 
Table~S1).
For monolayer ScI$_2$, the magnetic interaction is dominated by the nearest-neighbor (1NN) 
intralayer exchange between localized Sc moments. From the DFT calculations, the 1NN 
intralayer exchange constant is estimated to be $J_{\parallel}^{\mathrm{1NN}} = 33.2$ meV, 
indicating a strong ferromagnetic coupling within a monolayer. This relatively large energy 
scale is consistent with the narrow Sc-$d$-derived bands near the Fermi level and serves as the reference interaction strength for the bilayer systems. 
In the bilayer calculations, the intralayer exchange interaction is found to remain nearly 
unchanged, reflecting the weak influence of interlayer van der Waals bonding 
on the intrinsic magnetic coupling within each layer.

In contrast, the interlayer magnetic exchange interaction exhibits a pronounced dependence on the stacking configuration. By comparing total energies 
corresponding to ferromagnetic and antiferromagnetic alignments of the magnetic moments in the two layers, both nearest-neighbor (1NN) and second-
nearest-neighbor (2NN) interlayer exchange interactions were extracted. The resulting exchange constants for all stacking configurations are 
summarized in Table~\ref{tab:Mag_exchng}. For the AA-stacked bilayer, the 1NN interlayer exchange interaction is ferromagnetic with a value of 0.93 
meV, while the corresponding 2NN interaction amounts to 0.17 meV. In the AB-stacked configuration, both 1NN and 2NN interlayer exchange interactions 
become antiferromagnetic, with values of $-0.27$ meV and $-0.125$ meV, respectively. For the BA stacking, the interlayer exchange reverts to a 
ferromagnetic character, with 1NN and 2NN exchange constants of 0.16 meV and 0.083 meV, respectively.

As summarized in Table~\ref{tab:Mag_exchng}, these results clearly demonstrate that the sign 
and magnitude of the interlayer magnetic exchange in bilayer ScI$_2$ can be tuned by relative 
lateral displacement of the layers. The stacking-dependent reversal between ferromagnetic and 
antiferromagnetic interlayer coupling can be traced back to modifications in interlayer Sc--I-
-Sc exchange pathways, as discussed in the electronic structure section. Although the 
interlayer exchange interactions are an order of magnitude smaller than the intralayer 
coupling, they play a decisive role in determining the magnetic ground state and finite-
temperature magnetic ordering of the bilayer system.

%%%%%%%  Table  Magnetic Exchange %%%%%%%%%%
\begin{table}[h]
\caption{\label{tab:Mag_exchng}Interlayer magnetic exchange interactions, first nearest neighbor (1NN) and second nearest neighbor (2NN), for different configurations of \Sc.  Positive (negative) values indicate ferromagnetic (antiferromagnetic) coupling.}
\begin{ruledtabular}
\begin{tabular}{l|c c}
\multirow{2}{*}{Config.} & \multicolumn{2}{c}{Magnetic exchange interaction in meV}   \\
                  &    1NN  & 2NN \\
                  \hline
           AA     &    0.93   &    0.17 \\
           AB     &   -0.27   &   -0.13 \\
           BA     &    0.16   &    0.08 \\
\end{tabular}
\end{ruledtabular}
\end{table}
%%%%%%%% Section MAE %%%%%%%%
\subsection{Magnetic anisotropy energy}

While the stacking-dependent interlayer exchange interactions discussed in the previous section determine the relative alignment of magnetic moments 
across the layers, magnetic anisotropy plays a complementary role by stabilizing long-range magnetic order in low-dimensional systems. The presence 
of a finite magnetic anisotropy energy (MAE) lifts the continuous spin-rotational symmetry and suppresses long-wavelength spin fluctuations, thereby 
enabling finite-temperature magnetic ordering. In this context, we examine the MAE of monolayer and bilayer ScI$_2$ and its dependence on stacking 
configuration.

The MAE was evaluated by including spin--orbit coupling within the noncollinear density functional theory framework and computing the total energy difference between different magnetization orientations. Specifically, the MAE is defined as
\begin{equation}
\Delta E_\mathrm{MAE} = E_{110} - E_{001},
\end{equation}
where $E_{110}$ and $E_{001}$ denote the total energies with magnetization oriented along the 
in-plane $[110]$ and out-of-plane $[001]$ directions, respectively. Within this convention, a 
positive MAE corresponds to an in-plane magnetic easy axis.

For monolayer ScI$_2$, the calculated MAE is 0.07~eV per formula unit, indicating a well-
defined magnetic easy axis along the out-of-plane [001] direction. This relatively large 
anisotropy reflects the combined effects of reduced dimensionality and spin--orbit coupling, 
primarily originating from the heavy iodine atoms. Together with the strong intralayer 
ferromagnetic exchange interaction discussed earlier, this out-of-plane anisotropy supports 
the stabilization of long-range magnetic order in the monolayer system.

Importantly, the bilayer ScI$_2$ systems exhibit a qualitatively similar anisotropic behavior. For the AA, AB, and BA stacking configurations, the 
calculated MAE values are 0.05, 0.06, and 0.05~eV per formula unit, respectively. In all three cases, the MAE remains positive, indicating that the 
magnetic easy axis continues to lie along the out-of-plane [001] direction. The relatively small variation of MAE among different stacking 
configurations suggests that magnetic anisotropy in bilayer ScI$_2$ is only weakly affected by stacking geometry, in contrast to the pronounced 
stacking dependence observed for the interlayer exchange interactions.

The persistence of out-of-plane magnetic anisotropy from monolayer to bilayer ScI$_2$ indicates that interlayer coupling does not significantly alter 
the spin--orbit-induced anisotropic energy landscape. Although the magnitude of the MAE is slightly reduced in the bilayer compared to the monolayer, 
it remains sufficiently large to stabilize long-range magnetic order. The MAE values obtained here, together with the exchange parameters discussed 
in the previous section, provide the essential input parameters for the Monte Carlo simulations used to estimate the magnetic ordering temperatures, 
which are presented in the following section.

\subsection{Finite-temperature magnetic ordering and Monte Carlo simulations}

The magnetic exchange parameters and magnetic anisotropy energies obtained from first-principles calculations provide the microscopic basis for assessing the finite-temperature magnetic behavior of bilayer ScI$_2$. While density functional theory is inherently restricted to zero temperature, the stability of magnetic order against thermal fluctuations can be reliably examined using finite-temperature Monte Carlo simulations. In the present work, classical Monte Carlo simulations were carried out for bilayer ScI$_2$ in the AA, AB, and BA stacking configurations, using exchange parameters derived from first-principles calculations and incorporating the out-of-plane magnetic anisotropy discussed in the previous section.

\begin{figure*}[h]
    \centering
    \includegraphics[width=0.9\linewidth]{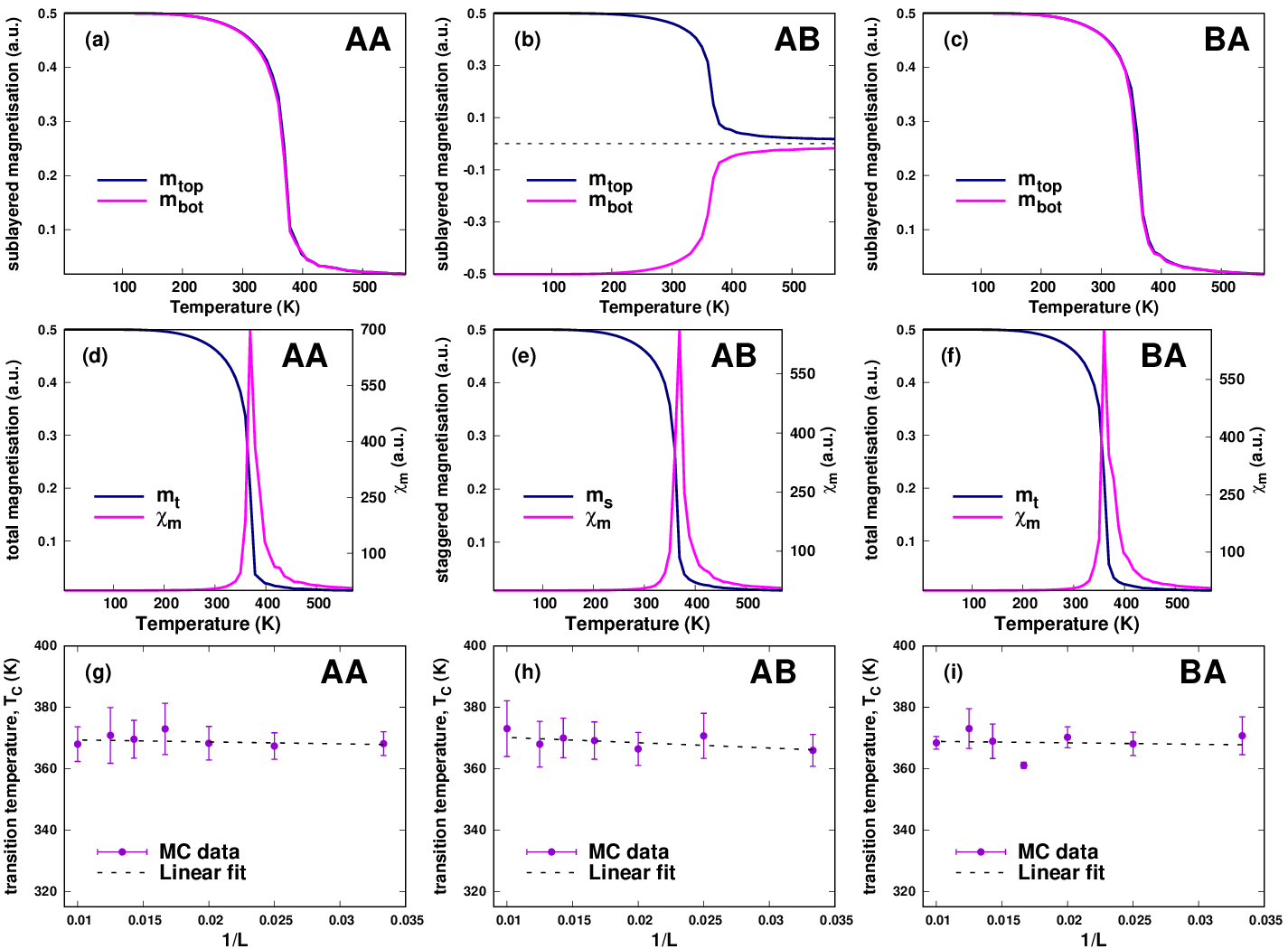}
    \caption{Finite-size scaling analysis of the magnetic transition temperature in bilayer ScI$_2$. Panels (a)–(c) show the temperature dependence of the top and bottom sublayered magnetisations, obtained on a square computational grid of dimensions $100\times100$, for the AA, AB, and BA stacking configurations, respectively. Panels (d)–(f) display the corresponding total magnetisation (for ferromagnetic: AA and BA) and staggered magnetisation (for antiferromagnetic: AB) and magnetic susceptibility as a function of temperature. Panels (g)–(i) present the weighted linear extrapolation of the pseudo-critical temperature $T_c$ as a function of $1/L$, where the y-intercept yields the transition temperature in the thermodynamic limit.}
    \label{fig:mc}
\end{figure*}

The temperature dependence of the magnetization and magnetic susceptibility was evaluated for several system sizes to characterize the nature of the magnetic phase transitions. For the AA- and BA-stacked bilayers, which exhibit ferromagnetic ground states, the magnetization decreases rapidly with increasing temperature and vanishes near a well-defined critical temperature, signaling a transition from an ordered ferromagnetic phase to a paramagnetic state. In the same temperature range, the magnetic susceptibility displays a pronounced peak, providing an initial estimate of the ordering temperature. These features are clearly visible in the temperature-dependent magnetization and susceptibility curves shown in Fig.~\ref{fig:mc}(a--f).

In contrast, the AB-stacked bilayer stabilizes an antiferromagnetic ground state, resulting 
in a strongly suppressed net magnetization over the entire temperature range due to the 
antiparallel alignment of spins in the two layers. Consequently, the thermal response of the 
AB stacking is not characterized by a finite total magnetization but is instead reflected 
through enhanced magnetic fluctuations and corresponding features in the susceptibility. This 
qualitative distinction between ferromagnetic and antiferromagnetic stackings is consistently 
captured by the Monte Carlo simulations and mirrors the magnetic ground states predicted from 
first-principles calculations.

A more accurate and size-independent determination of the magnetic ordering temperature is 
obtained from the analysis of the fourth-order Binder cumulant calculated for multiple system 
sizes. The crossing points of the Binder cumulant curves within a narrow temperature window 
provide a robust estimate of the critical temperature, as shown in Fig.~\ref{fig:mc}(g--i). 
This analysis confirms the presence of long-range ferromagnetic order in the AA and BA 
stackings and the absence of a net magnetization in the antiferromagnetic AB configuration, 
while still allowing a well-defined thermal transition to be identified.

The critical temperatures extracted from the Binder cumulant crossings are found to lie in 
the range of approximately 360--375~K for all three stacking configurations and are 
summarized in Table~\ref{tab:Tc}. Notably, despite the distinct magnetic ground states and 
stacking-dependent interlayer exchange interactions, the ordering temperatures remain 
remarkably similar across all configurations. This observation indicates that the dominant 
energy scale governing the thermal stability of magnetic order is the strong intralayer 
exchange interaction, whereas the stacking-dependent interlayer coupling primarily determines 
the nature of the magnetic ground state rather than the magnitude of the transition 
temperature.

To estimate the transition temperature in the thermodynamic limit, a finite-size scaling analysis was performed using the Binder-cumulant-derived critical temperatures for different system sizes. The thermodynamic-limit transition temperature $T_c(\infty)$ was obtained by fitting the finite-size estimates $T_c(L)$ using the scaling relation
\begin{equation}
T_c(L) = T_c(\infty) + a L^{-1/\nu},
\end{equation}
where $L$ is the linear system size and $\nu$ is the correlation-length exponent\cite{Diaz2017PhysA}. Given the quasi-two-dimensional Ising-like nature of the system enforced by the strong out-of-plane magnetic anisotropy, the exponent was taken as $\nu = 1$, leading to a linear dependence on $1/L$. A weighted linear fit, with weights determined by the statistical uncertainties of the Binder cumulant crossings, yields a reliable estimate of the transition temperature in the thermodynamic limit. Such a statistical estimation of $T_{c}$ is indeed reliable in 2D magnetic systems \cite{Chandra2021JPCS}.

Overall, the Monte Carlo simulations demonstrate that bilayer ScI$_2$ sustains robust magnetic order at and above room temperature for all considered stacking configurations. At the same time, the magnetic ground state---ferromagnetic or antiferromagnetic---can be selectively tuned through stacking geometry without significantly affecting the ordering temperature. This combination of high thermal stability and stacking-controlled magnetic order underscores the potential of bilayer ScI$_2$ as a model platform for exploring tunable magnetism in two-dimensional van der Waals materials.

%%%%%%%%%%  Figure Monte Carlo Simulations  %%%%%%%%

%%%%%%%%%%  Table Monte Carlo %%%%%%%%%%
\begin{table}[t]
	\centering
	\caption{Critical temperatures $T_c$ obtained from Binder cumulant crossings for different stacking configurations of bilayered $ScI_{2}$. The corresponding magnetic ground states are also indicated.}
	\label{tab:Tc}
	\begin{ruledtabular}
		\begin{tabular}{lcc}
			Stacking & Magnetic order & $T_c$ (K) \\
			\hline
			$AA$ & Ferromagnetic & $370.0 \pm 5.8$ \\
			$AB$ & Antiferromagnetic & $371.9 \pm 6.8$ \\
			$BA$ & Ferromagnetic & $369.4 \pm 3.5$ \\
		\end{tabular}
	\end{ruledtabular}
\end{table}

\subsection*{}
%\subsection*{Generality of stacking-controlled interlayer magnetic exchange}
The stacking-dependent interlayer magnetic coupling observed in bilayer \Sc\ reflects a general mechanism operative in layered magnetic 
materials, where subtle changes in atomic registry can qualitatively modify the superexchange pathways and consequently alter the 
magnetic ground state. Similar stacking-controlled magnetic phase transitions have been reported in several two-dimensional van der 
Waals magnets, particularly in bilayer CrI$_3$ and related transition-metal halides, where lateral shifts between adjacent layers induce 
reversible switching between FM and AFM interlayer coupling 
\cite{Huang2017Nature, burch2018magnetism,Sivadas2018NanoLett,Gong2017Discovery}. 
Recent theoretical studies on bilayer ScI$_2$ have also 
demonstrated that stacking geometry strongly influences the interlayer magnetic ordering and can further induce multiferroic behavior 
through interlayer sliding and symmetry breaking \cite{Pan2025ApplPhysLett}. 
Our present results are fully consistent with these reports 
and further provide a detailed microscopic understanding of how orbital hybridization and exchange pathways govern this behavior.\\
In the AA and BA stacking configurations, the relative alignment of the layers favors interlayer superexchange pathways characterized by 
near-orthogonal overlap between the dominant Sc-$d$ orbitals and I-$p$ orbitals across the interface. Within the Goodenough--Kanamori--
Anderson framework, such geometries suppress antiferromagnetic kinetic exchange and allow ferromagnetic superexchange contributions, 
often stabilized by Hund’s coupling on the ligand $p$ orbitals \cite{Goodenough1963,Kanamori1959,Anderson1959}. As a result, the 
effective interlayer exchange interaction remains ferromagnetic in these stackings. This mechanism does not rely on direct interlayer 
metal--metal overlap but arises from the symmetry of the metal--ligand--metal exchange paths, making it broadly applicable to van der 
Waals magnets with similar bonding motifs.\\
In contrast, the AB stacking introduces a registry in which the relative lateral displacement of the layers enhances symmetry-allowed 
overlap between compatible Sc-$d$ and I-$p$ orbitals across the interface. This configuration promotes virtual hopping processes that 
favor antiparallel spin alignment, thereby stabilizing antiferromagnetic superexchange in accordance with the Goodenough--Kanamori--
Anderson rules \cite{Goodenough1963,Anderson1959}. Compared with previous reports that primarily identified the FM--AFM switching 
behavior, our analysis further emphasizes the specific role of first- and second-nearest-neighbor exchange pathways and their dependence 
on stacking geometry, providing a clearer physical picture of the microscopic origin of the magnetic transition.\\
From a broader perspective, the coexistence of stacking-sensitive interlayer exchange and stacking-insensitive intralayer magnetism 
highlights an important design principle for two-dimensional magnetic materials. While strong intralayer exchange interactions determine 
the dominant energy scale and largely control the magnetic ordering temperature, stacking geometry offers an efficient and reversible 
route for tuning the interlayer magnetic ground state. This separation of energy scales explains why stacking can induce transitions 
between FM and AFM interlayer order without significantly reducing the thermal stability of magnetism. Such behavior not only agrees 
with previous theoretical predictions for ScI$_2$ but also demonstrates the potential of stacking engineering as a practical strategy 
for designing controllable magnetic states in layered spintronic devices.
\section{Conclusions}
In summary, we have carried out a comprehensive first-principles and finite-temperature study of stacking-dependent magnetic ordering 
in bilayer \Sc. Using density functional theory combined with Hubbard-$U$ corrections, phonon stability analysis, and Monte Carlo 
simulations, we systematically investigated the structural, electronic, and magnetic properties of monolayer and bilayer ScI$_2$ in 
three distinct stacking configurations. The calculated phonon spectra confirm the dynamical stability of the considered bilayer 
structures, while the electronic structure reveals a spin-polarized ground state dominated by Sc-$d$ states near the Fermi level, 
providing the microscopic origin of magnetism in this system.\\
By mapping total energies onto an effective Heisenberg spin Hamiltonian, we show that although the strong intralayer ferromagnetic 
exchange remains nearly unaffected by stacking, the interlayer magnetic exchange is highly sensitive to the relative stacking geometry. 
Specifically, AA and BA stackings favor ferromagnetic interlayer coupling, whereas AB stacking stabilizes an antiferromagnetic 
interlayer alignment. This FM--AFM switching originates from stacking-induced modifications of the Sc--I--Sc superexchange pathways and 
orbital hybridization. Magnetic anisotropy calculations further reveal a robust out-of-plane easy axis, ensuring the stability of long-
range magnetic order in this quasi-two-dimensional system.\\
Finite-temperature Monte Carlo simulations based on the DFT-derived exchange parameters and anisotropy energies reveal that bilayer 
\Sc~sustains magnetic ordering at and above room temperature for all considered stacking configurations. Despite the stacking-
dependent magnetic ground states, the ordering temperatures remain within a narrow range, highlighting the dominant role of intralayer 
exchange in controlling thermal stability, while stacking primarily tunes the interlayer magnetic alignment.\\
The combination of high magnetic ordering temperature, stacking-tunable interlayer magnetism, robust magnetic anisotropy, and dynamical 
stability makes bilayer \Sc~a promising platform for controllable magnetism in two-dimensional van der Waals materials. The ability 
to reversibly switch between ferromagnetic and antiferromagnetic interlayer coupling through stacking geometry may be exploited in 
spintronic and magneto-electronic devices, such as spin-filtering junctions and stacking-engineered magnetic heterostructures. More 
broadly, our results establish stacking engineering as an effective strategy for designing tunable magnetic states in layered magnets.

\begin{acknowledgments}
This work was supported by the Anusandhan National Research Foundation (formerly Science and Engineering Research Board) 
for the computational support through the start-up research grant (SRG/2023/000122).
\end{acknowledgments}

\section*{Data Availability Statement}
Data sets generated during the current study are available from the corresponding author on reasonable request.

\bibliography{myRef}% Produces the bibliography via BibTeX.
\end{document}